**Integrating diverse datasets improves developmental enhancer prediction**

Genevieve D. Erwin[1], Rebecca M. Truty[1], Dennis Kostka[2], Katherine S. Pollard[1,3], John A. Capra[4]

[1] Gladstone Institute of Cardiovascular Disease
[2] Department of Developmental Biology, and Department of Computational and Systems Biology, University of Pittsburgh, Pittsburgh, PA 15201, USA
[3] Department of Epidemiology and Biostatistics, and Institute for Human Genetics, University of California, San Francisco, CA 94158, USA
[4] Center for Human Genetics Research, Department of Biomedical Informatics, Vanderbilt University, Nashville, TN 37232, USA

**Corresponding authors**
Katherine S. Pollard
Gladstone Institute of Cardiovascular Disease
1650 Owens Street
San Francisco, CA 94158
(415) 734-2000
Email: kpollard@gladstone.ucsf.edu

John A. Capra
Center for Human Genetics Research
519-A Light Hall
2215 Garland Ave
Nashville, TN 37232
(615) 343-3671
Email: tony.capra@vanderbilt.edu

**Running title:** Integrative developmental enhancer prediction

## ABSTRACT

Gene-regulatory enhancers have been identified by many complementary lines of evidence, including evolutionary conservation, regulatory protein binding, chromatin modifications, and DNA sequence motifs. To integrate these different approaches, we developed EnhancerFinder, a novel method for predicting developmental enhancers and their tissue specificity. EnhancerFinder uses a two-step multiple-kernel learning approach to integrate DNA sequence motifs, evolutionary patterns, and thousands of diverse functional genomics datasets from a variety of cell types and developmental stages. We trained EnhancerFinder on hundreds of experimentally verified human developmental enhancers from the VISTA Enhancer Browser—in contrast to histone mark or sequence-based enhancer definitions commonly used. We comprehensively evaluated EnhancerFinder, and we found that our integrative approach improves enhancer prediction accuracy over previous approaches that consider a single type of data, such as evolutionary conservation and the binding of enhancer-associated proteins. Our tissue-specific prediction evaluation underscores the importance of considering enhancers of different tissues, in addition to non-enhancer regions, when attempting to learn specific types of enhancers. We find that VISTA enhancers active in embryonic heart are easier to predict than enhancers active in several other embryonic tissues due to their uniquely high GC content. We applied EnhancerFinder to the entire human genome and predicted 84,301 developmental enhancers and their tissue specificity. These predictions provide specific functional annotations for large amounts of human non-coding DNA, and are significantly enriched near genes with annotated roles in their predicted tissues and hits from genome-wide association studies. We demonstrate the utility of our enhancer predictions by identifying and validating a novel cranial nerve enhancer in the *ZEB2* locus. Our genome-wide developmental enhancer predictions are freely available as a UCSC Genome Browser track.

## AUTHOR SUMMARY

The human genome contains an immense amount of non-coding DNA with unknown function. Some of this DNA is responsible for regulating exactly when and where genes are turned on during development. Enhancers, which are a type of regulatory element, are short stretches of DNA that can act as "switches" to turn a gene on or off at specific times in specific cells or tissues. Understanding where in the genome enhancers sit can provide insight into development, genetic diseases, and even response to drug treatment. Enhancers are hard to find, but clues to their locations are found in different types of data including DNA sequence, certain proteins' binding to the DNA, and the evolutionary history of that stretch of DNA. We introduce a new method, called EnhancerFinder, that combines thousands of newly available datasets to predict the location and activity of enhancers. EnhancerFinder was trained on the largest set of known human enhancers available, and we find that it works very well. We use EnhancerFinder to predict tens of thousands of enhancers in the human genome. These predictions will be useful in understanding functional regions hidden in the vast amounts of non-coding DNA.

## INTRODUCTION

Eukaryotic gene expression is regulated by a highly orchestrated network of events, including the binding of regulatory proteins to DNA, chemical modifications to DNA and nucleosomes, recruitment of the transcriptional machinery, splicing, and post-transcriptional modifications. Enhancers are genomic regions that influence the timing, amplitude, and tissue specificity of gene expression through the binding of transcription factors and co-factors that increase transcription (as reviewed in [1,2]). In humans, genetic variation in enhancer regions is implicated in a wide variety of developmental disorders, diseases, and adverse responses to treatments [3,4].

Enhancers have been discovered in introns, exons, intergenic regions megabases away from their target genes [5], and even on different chromosomes [6]. An enhancer frequently drives only one of many domains of a gene's expression [7,8] and different cell types accordingly exhibit considerable differences in their active enhancers [9,10]. This modularity enables the creation of complex regulatory programs that can evolve relatively easily between closely related species [11,12].

Individual enhancers were initially identified using transgenic assays in cultured cell lines [13,14] and later *in vivo* in model organisms, such as mouse, *Drosophila*, and zebrafish. In the *in vivo* experiments, a construct containing the sequence to be tested for enhancer activity, a minimal promoter, and a reporter gene (e.g., lacZ) is injected into fertilized eggs, and transgenic individuals are assayed for reporter gene expression. Studies of this type typically investigate regulation of a single locus, as has been done for *Bmp5*, *Shh*, and *RET* [15,16,17], because transgenic assays are too low throughput for genome-wide analyses.

Early efforts to find enhancers at the genome scale used comparative genomics. Several studies assayed non-coding regions conserved across diverse species for enhancer activity [18,19,20], since functional non-coding regions likely evolve under negative selection. This approach identified many enhancers at a range of levels of evolutionary conservation [21,22,23]. However, relying on evolutionary conservation alone has several shortcomings: many characterized enhancers are not conserved between species [24], non-coding conservation is not specific to enhancer elements, and evolutionary patterns provide little information about the tissue and timing of enhancer activity.

Enhancer prediction has been revolutionized by recent technological advances, including chromatin immunoprecipitation coupled with high-throughput sequencing (ChIP-seq) [25], RNA sequencing (RNA-seq), and sequencing of DNaseI-digested chromatin (DNase-seq) [26] or formaldehyde-assisted isolation of regulatory elements (FAIRE-seq) [27]. These "functional genomics" assays enable genome-wide measurement of histone modifications, binding sites of regulatory proteins, transcription levels, and the structural conformation of DNA. The ENCODE project [28] and similar studies focused on specific developmental cell types [29,30] have dramatically increased the amount of publicly available functional genomics data.

Functional genomics studies revealed several genomic signatures of active enhancers. For example, known enhancers are associated with the unstable histone variants H3.3 and H2A.Z [31,32] and low nucleosome occupancy [33], although these chromatin states are not unique to enhancers. Monomethylation of lysine 4 on histone H3 (H3K4me1), a lack of trimethylation at the same site (H3K4me3), and acetylation of lysine 27 on histone H3 (H3K27ac) may distinguish active enhancers from promoters [9,34,35], enhancers that are "poised" for activity later in development [36,37], and regulatory elements that repress gene expression [38,39]. Additional features that pinpoint specific classes of active enhancers include binding of the transcriptional cofactor p300/CBP [20,40,41,42], clusters of transcription factor (TF) binding sites [43,44,45,46], and enhancer RNA transcription (eRNAs)[47]. Collectively, functional genomics data have pinpointed the locations of many novel enhancers and yielded insights into sequence and structural determinants of enhancer activity. However, these patterns have not proven to be universal [48,49], and there is unlikely to be a single chromatin signature that identifies all classes of enhancers [10,50,51].

Given the complexity of these functional genomics data sets, computational methods have been developed to improve and generalize the enhancer predictions made from simple combinations of these data. Support vector machines (SVMs) and linear regression models trained to interpret DNA sequence motifs underlying known enhancers have successfully identified novel enhancers active in heart [52], hindbrain [53], and muscle [54] development. Another approach used SVMs to learn patterns of short DNA sequence motifs that distinguish markers of potential enhancers, such as p300 and H3K4me1, in different cellular contexts [55,56]. Random forests have been used to predict p300 binding sites from histone modifications in human embryonic stem cells and lung fibroblasts [57]. Machine-learning algorithms have also been applied to the related problem of selecting functional TF binding sites out of the thousands of hits to a TF's binding motif throughout the genome [58,59,60,61,62,63,64]. Finally, two groups have taken a less supervised approach and used hidden Markov models (ChromHMM) [65] and dynamic Bayesian networks (Segway) [66] to segment the human genome into regions with unique signatures in ENCODE data and to assign potential functions, such as enhancer activity, to these states.

While rich datasets coupled with sophisticated algorithms have successfully identified many novel enhancers, comprehensive enhancer prediction is challenging for three main reasons. First, no single type of data is currently sufficient to identify all enhancers active in a given context. Many of the approaches described above use a single mark or motif as a proxy for an enhancer, but this gives an incomplete representation of all biologically active enhancers. Second, while a great deal of functional genomics data is available for different cell lines and tissues, it is not understood how informative experiments in a given cellular context are about enhancer activity in other contexts. Finally, many different approaches have been taken to find enhancers using functional genomics data and/or sequence features, and the relative merits of each technique are not well understood.

With these issues in mind, we introduce a new two-step machine-learning method for predicting enhancers and their tissue specificity. In machine learning, a classification algorithm is trained to distinguish between labeled training examples (e.g., enhancers and non-enhancers) based on features of these labeled examples (e.g., evolutionary conservation, chromatin signature, DNA sequence). The trained classifier can then be used to predict the labels for uncharacterized genomic regions (e.g., which ones are enhancers). Our approach employs two rounds of a supervised machine-learning technique called multiple kernel learning (MKL) [67,68]. MKL is based on the theory of SVMs [69], but provides greater flexibility to combine diverse data (e.g., evolutionary conservation, sequence motifs, and functional genomics data from different cellular contexts) and to interpret their relative contributions to the resulting predictions. In our two-step MKL algorithm, which we call EnhancerFinder, we first train a classifier to distinguish known human developmental enhancers in the VISTA enhancer database [70] from the genomic background. In the second step, we train classifiers to distinguish among enhancers in VISTA with different tissue specificity. In contrast to most other enhancer prediction strategies, which are trained on a epigenetic marks or sequence motif that serve as a proxy for a subset of all active enhancers, our use of a heterogeneous and *in vivo* validated set of enhancers, enables us to characterize the complex suite of features that underlie active regulatory regions. Our two-step approach allows us to determine more accurate signatures of enhancers with activity in specific contexts.

Our analyses demonstrate that EnhancerFinder's integration of diverse types of data from many different cellular contexts significantly improves prediction of validated enhancers over previous approaches. We find that enhancers active in some developmental contexts are easier to identify than others. Applying EnhancerFinder to the entire human genome, we predict more than 80,000 developmental enhancers, with tissue-specific predictions for brain, limb, and heart. These predictions significantly overlap known non-coding regulatory regions and are enriched near relevant genome-wide association study hits and genes expressed in the predicted tissue. To illustrate the utility of our enhancer predictions, we use them to investigate the regulation of the *ZEB2* gene and experimentally validate a new cranial nerve enhancer nearby.

**RESULTS**

We present EnhancerFinder, a novel machine learning-based enhancer prediction pipeline that allows the seamless integration of feature data from a variety of experimental techniques and biological contexts that have previously been used individually to predict enhancers (Figure 1). We use MKL to integrate these data. MKL algorithms learn a weighted combination of different "kernel" functions that quantify the similarity of different feature data in order to make predictions. In EnhancerFinder, we use three kernels based on different types of biological feature data: DNA sequence motifs, evolutionary conservation patterns, and 2496 functional genomics datasets.

EnhancerFinder could be used to predict enhancers active in any life stage and tissue. In this study, we evaluate EnhancerFinder's ability to predict developmental enhancers and their tissue specificity. Step 1 of our pipeline aims to distinguish all enhancers active in the context of interest (i.e., a specific developmental stage) from non-enhancer regions. Step 2 then builds classifiers to predict the tissues in which the enhancer candidates from Step 1 are active. In contrast to one-step approaches, this two-step approach allows us to accurately identify enhancers, while also distinguishing their tissues of activity.

We train and evaluate EnhancerFinder using the VISTA Enhancer Browser, which contains over 700 human sequences with experimentally validated enhancer activity in at least one tissue at embryonic day 11.5 (E11.5) in transgenic mouse embryos. VISTA also contains a similar number of regions without enhancer activity in this context. E11.5 in mouse development roughly corresponds to E41 (Carnegie stage 17 [71]) in human development. In Step 1 of EnhancerFinder, we used all 711 VISTA enhancers as positive training data, and for negative training data, we created a set of 711 random regions matched to the length and chromosome distribution of the positives to represent the genomic background. In Step 2, we considered all enhancers in VISTA with activity in a tissue of interest as positives and all other regions in VISTA (including enhancers of other tissues and regions not found to be active at E11.5) as negatives (see Methods). This second step that includes enhancers active in other tissues in the training proves to be essential for obtaining high specificity in predicting tissue of activity.

To evaluate EnhancerFinder, we compared it to several commonly used enhancer prediction approaches. We evaluated the performance of all prediction algorithms using 10-fold cross validation to compute the area under the curve (AUC) for receiver operating characteristic (ROC) curves. We also computed precision-recall curves (Supplementary Figure 1) and compared power at a low false positive rate.

**EnhancerFinder integrates diverse data types to accurately identify developmental enhancers**

EnhancerFinder predicts enhancers by integrating classifiers based on three distinct data types: functional genomics data, evolutionary conservation patterns, and DNA sequence motifs. Combining these different approaches enables EnhancerFinder to accurately distinguish enhancers from the genomic background (Figure 2A; AUC = 0.96).

The functional genomics component of EnhancerFinder (which we refer to as **All Functional Genomics**) is a linear SVM that incorporates 2469 datasets generated by the ENCODE project and smaller scale studies. These include DNaseI hypersensitivity data and ChIP-Seq for p300, many histone modifications, and many TFs from many adult and embryonic tissues and cell lines (Supplementary Table 1). DNA sequence patterns are integrated via a 4-spectrum kernel (**DNA Motifs**), which summarizes the occurrence of all length four DNA sequences (4-mers) in input regions [72]. We found that little was gained by increasing $k$, considering multiple $k$ simultaneously, or incorporating knowledge of transcription factor binding site (TFBS) motifs (Supplementary Figures 2 and 3). Finally, evolutionary conservation information is incorporated with a linear SVM that uses mammalian phastCons scores [73] as features (**Evolutionary Conservation**).

**EnhancerFinder performs significantly better than previous enhancer prediction approaches**

Our motivation for developing EnhancerFinder was to explore whether combining previous successful approaches to enhancer prediction would improve performance. Each of the classifiers combined in EnhancerFinder is representative of a different strategy for predicting enhancers. Thus, we compared the performance of EnhancerFinder to each of its constituents, which are SVMs trained on the same enhancer data as EnhancerFinder, but using only one type of the data features (e.g., only sequence motifs). EnhancerFinder significantly outperformed each of the individual classifiers (Figure 2A; p=2.0E-7 for **Evolutionary Conservation**, p=2.6E-8 for **DNA Motifs**, and p=4.4E-16 for **All Functional Genomics**, McNemar's test), suggesting that these different types of data capture unique aspects of enhancers that are not completely encompassed by any single data type. Of the three component classifiers in EnhancerFinder, **Evolutionary Conservation** yields the best performance (AUC=0.93). This reflects the fact that nearly all validated enhancers in VISTA are strongly conserved compared to the genomic background and many were selected for testing based on their conservation. This is surely an over-estimate of the performance of this approach, and we address this issue in more detail when describing our tissue-specific and genome-wide predictions below. The **DNA Motifs** (AUC=0.88) and **All Functional Genomics** (AUC=0.89) classifiers also exhibit strong performance, but do not perform as well as the combined classifier. EnhancerFinder has nearly twice the power of any of the individual classifiers at a 5% false positive rate (FPR), and its power advantage is even larger at lower FPRs.

**All Functional Genomics**, **DNA Motifs**, and **Evolutionary Conservation** achieve roughly similar performance from different feature data, but each individual classifier predicts a somewhat different set of enhancers during evaluation (Figure 2B). Roughly two-thirds of the enhancer predictions are shared between the three classifiers. The improvement provided by combining these data argues that these data sources are indeed complementary.

We also compared EnhancerFinder's performance with CLARE, a popular method for identifying enhancers from DNA sequence data, i.e., transcription factor binding site motifs and other sequence patterns [74]. This approach has been successfully applied in several contexts [52,53,54,75] and is publicly available as a web server. On our Step 1 enhancer prediction task, we find that CLARE achieves an ROC AUC of 0.79. This is much lower than our **DNA Motifs** method (AUC=0.88) and the full **EnhancerFinder** (AUC=0.96; Figure 2C). At a 5% FPR, the power of CLARE is about 20%, compared to approximately 30% for **DNA Motifs** and more than 60% for **EnhancerFinder**.

Finally, we compared EnhancerFinder to ChromHMM and Segway [65,66], two unsupervised machine learning methods for segmenting the genome into a small number of functional "states" based on consistent patterns in ENCODE data. The states resulting from the segmentations were annotated by hand into predicted functional classes, which include enhancer activity. It is difficult to directly compare our supervised method to these unsupervised approaches, but we believe it is informative, since they are commonly used to interpret uncharacterized non-coding regions. To evaluate these methods, we considered the states overlapping our training and testing regions. Any region with an overlapping enhancer state was considered a predicted enhancer and all others were predicted non-enhancers. In this way, we obtained a single point in performance evaluation space for the state predictions. Since there is no score or confidence value associated the state assignments, a full ROC curve could not be created for these methods. Figure 2C gives the performance for several versions of ChromHMM and Segway based on ENCODE data from different cell lines. Both methods perform better than random, but considerably worse than other approaches (p≈0). We stress that, in contrast to our supervised method, these methods were not explicitly trained to perform the same task as EnhancerFinder, and instead are attempting to solve a more general problem. However, this result argues that their utility in identifying developmental enhancers is limited compared to specialized approaches.

**Integrating diverse functional genomics data improves enhancer prediction**
As illustrated above, our machine learning prediction and evaluation framework enabled us to quantitatively explore the utility of different genomics datasets in enhancer prediction by creating classifiers based on types of data (i.e., sequence motifs, evolutionary conservation, and functional genomics) and comparing their performance. We also used this framework to investigate other questions about the utility of different subsets of these data for enhancer prediction. For example, one might expect that some of the datasets included in **All Functional Genomics** (e.g., experiments in cancer cell lines or adult tissues) would not be as useful as others (e.g., experiments in embryonic tissues) for predicting developmental enhancers, and that limiting the features examined by the classifier to the most relevant experiments might improve performance.

To explore this hypothesis, we trained linear SVM classifiers to predict VISTA enhancers (as in Step 1 of EnhancerFinder) based on different subsets of all the functional genomics features (Supplementary Table 1) and compared their performance. First, we considered a collection of 244 datasets from embryonic tissues and cell lines (**Embryonic Functional Genomics**). Next, we created a classifier that considers data from a wider range of contexts by training a linear SVM using a large, manually curated set of 509 potentially relevant functional genomics data sets (**Relevant Functional Genomics**). This set includes embryonic datasets, along with additional DNaseI and ChIP-Seq data from adult tissues and cell lines related to the dominant tissues of activity in VISTA. For example, we included data from human cardiac myocytes, since there are many developmental heart enhancers in our training examples. We compared these to the **All Functional Genomics** classifier described above that uses all 2496 functional genomics features.

**All Functional Genomics** (AUC=0.89) performed slightly better than **Relevant Functional Genomics** (AUC=0.87; p=0.16), and both significantly outperformed

**Embryonic Functional Genomics** (AUC=0.83; p=9.2E-9 and p=2.7E-6, respectively) (Figure 3A). At low FPRs, the differences in power between these classifiers were modest. The **Embryonic Functional Genomics** classifier included the most time-appropriate datasets, yet its performance was improved by including additional data sets that seem less relevant to our classification problem *a priori*. Thus, we conclude that it can be advantageous to consider a wide range of functional genomics features, many of which may not be directly associated with enhancer activity or measured in the context of interest. The utility of these additional data sets might indicate that some enhancer features are stable across cell types and developmental stages, but it could also reflect information these data provide about genomic regions that are *not* active enhancers during development (see Discussion).

**Histone marks and p300 provide complementary information about enhancer activity**

We also explored the utility of individual functional genomics datasets that are often used as proxies for developmental enhancers by creating three linear SVM classifiers: **H3K27ac**, **H3K4me1**, and **p300**. These SVMs were trained to distinguish VISTA positives from the genomic background (Step 1) using all available data of the specified type from ENCODE, which include a range of cell types and tissues (Supplementary Table 1). All three classifiers performed better than random (Figure 3B). **H3K4me1** (AUC=0.72) and **p300** (AUC=0.68) performed similarly (p=0.25), with **p300** performing best at low FPRs and **H3K4me1** best at higher FPRs. Both significantly outperformed **H3K27ac** (AUC=0.61; p=9.4E-15 and p=5.5E-9, respectively). Since combinations of these features are often used to predict enhancers, we next trained a linear SVM classifier (**Basic Functional Genomics**) that includes all three data types together. The combined classifier significantly outperforms all the individual classifiers (AUC=0.77; p<2E-7 for each), suggesting that each data type contributes unique information about enhancer activity. Also, all four SVM classifiers achieved much better performance than the common approach of simply considering regions overlapping with these data (Supplementary Figure 4).

EnhancerFinder also learns weights for individual features within classifiers that reflect their contribution to the enhancer predictions. We found that features known to be associated with enhancer activity in relevant cellular contexts generally receive positive weights, while those associated with other types of elements received negative weights (Supplementary Text and Supplementary Figure 5).

**EnhancerFinder's two-step approach enables tissue-specific enhancer prediction**

In the previous sections, we focused on generic developmental enhancer prediction (Step 1 of EnhancerFinder). Step 2 of EnhancerFinder applies a second round of MKL to refine and further annotate predicted enhancers from Step 1 (Figure 1). In this study, Step 2 consists of training an MKL classifier to distinguish VISTA enhancers active in a given tissue from VISTA regions without activity in that tissue, i.e., non-enhancers plus enhancers for other tissues. We did not require that the positive training examples be active *only* in the tissue of interest. Using the same feature data as in Step 1, we created tissue-specific classifiers for all tissues with more than 50 examples in VISTA: forebrain, midbrain, hindbrain, heart, limb, and neural tube.

The performance of EnhancerFinder's tissue specificity predictions varied dramatically between tissues (Figure 4), with the best performance for heart (AUC=0.85), followed by limb (AUC=0.74), forebrain (AUC=0.72), midbrain (AUC=0.72), hindbrain (AUC=0.69), and neural tube (AUC=0.62), which was the worst of the tested tissue classifiers, but better than random. We combined all brain enhancers into one class, and the performance of this generic brain classifier was similar to that of the more specific brain classifiers (AUC=0.73). Consistent with our observations from Step 1, the EnhancerFinder tissue-specific classifiers trained with all data types achieved the best performance for most tissues (Supplementary Table 2). However, classifiers based on functional genomics alone often performed almost as well as the full EnhancerFinder classifier, suggesting functional genomics data may be more informative about enhancer tissue specificity than conservation or sequence motifs.

Most previous efforts to predict tissue-specific enhancers have performed a single training step using enhancers or enhancer marks present in the tissue of interest as positives and non-enhancer regions or the genomic background as negatives. To test whether our two-step method improves upon these previous approaches, we trained one-step MKL classifiers and compared their predicted tissue distributions to those of validated enhancers from the VISTA database (Figure 5A). First, we trained a set of tissue-specific classifiers using enhancers active in a given tissue as positives and the genomic background as negatives. These classifiers predict very similar sets of enhancers regardless of the target tissue; and they vastly overestimate the number of VISTA enhancers that are active in multiple tissues (95% versus 8% of VISTA) and the number of true enhancers of each tissue (Figure 5B). In contrast, classifiers trained as in Step 2 of EnhancerFinder, i.e., using tissue-specific enhancers as positives and a mix of enhancers active in other tissues and regions with no VISTA activity as negatives, show much greater tissue-specificity in their predictions (76%) and a similar amount of overlap as in the known enhancers (Figure 5C).

**Heart enhancers are uniquely easy to identify due to their high GC content**
The relative ease of identifying heart enhancers is due in large part to their having several unique characteristics. Known heart enhancers at E11.5 are significantly less evolutionarily conserved than enhancers in other tissues [40,42]. In addition, we observed that heart enhancers at this developmental stage are uniquely close to the nearest transcription start site (TSS) (Supplementary Figure 6). However, the distinct attribute of heart enhancers most relevant to enhancer prediction is their unusually high GC content (49%). A simple classifier based solely on the GC content of a region performs nearly as well as our full classifiers (Supplementary Figure 7). In contrast, enhancers of other tissues at E11.5 exhibit much lower GC content (~40%) that is not significantly different from the genomic background, and as a result sequence-based classifiers do not perform well on the other tissues (Supplementary Table 2). The high GC content of heart enhancers is not due to overlap with CpG islands. Only about 4% of VISTA enhancers overlap with a CpG island, and this number is consistent across tissues. We do see, however, that repeat regions in heart enhancers are depleted for the very AT-rich repeats seen in other enhancers, and that most of the repeat regions in heart enhancers are 40-60% GC.

**We predict more than 80,000 developmental enhancers across the human genome**
One of the main motivations for developing algorithms that can distinguish active enhancers is to apply them to uncharacterized genomic regions to aid the exploration and interpretation of the gene regulatory landscape of the human genome (Figure 1). To produce a genome-wide set of candidate developmental enhancers, we divided the genome into 1.5 kb blocks overlapping one another by 500 bp and applied Step 1 of EnhancerFinder to each of these regions. EnhancerFinder produces a score for each region; positive scores indicate membership in the positive set (enhancers), and negative scores indicate membership in the negative set (non-enhancers). To focus on high confidence predictions in this genome-wide analysis, we used the cross-validation-based evaluation described above to find a 5% FPR score threshold, and only considered regions exceeding this threshold. After merging overlapping positive predictions, we identified 84,301 developmental enhancers across the human genome with median length of 1,500 bp and total genome coverage of 183,695,500 bp (5.86%).

The 5% FPR threshold we used corresponds to a 65% true positive rate (TPR). To calculate the false discovery rate (FDR), we must estimate the unknown fraction of 1.5 kb blocks of the human genome that harbor developmental enhancer regions. If this fraction were as high as 50%, a 5% FPR would correspond to a 9% FDR. If instead we estimate that 10% of 1.5 kb windows contain a developmental enhancer, we see an FDR of 47% at a 5% FPR. While this may seem high, our recent analysis of predicted enhancers with human-specific substitution rate acceleration found a much lower false validation rate at E11.5 (5/29) [75]. This suggests that these FDR estimates may be pessimistic, especially when predicted enhancer regions are analyzed in the context of other relevant data.

In our genome-wide analysis, we used the smaller **Relevant Functional Genomics** data set in order to reduce the computational time required. We also did not include evolutionary conservation data, because the positives in our training data are almost universally conserved; 706 of the 711 VISTA enhancers overlap a conserved element. While most enhancers likely exhibit some evolutionary conservation, this extremely high fraction is likely due to bias in the selection of the tested regions in VISTA and could reduce our ability to detect less highly conserved novel enhancers genome-wide (see the Discussion). The resulting conservation-free classifier still performed extremely well in cross validation (AUC=0.92). Supporting this approach, non-conserved regions make up over 20% of our genome-wide enhancer predictions. We did not observe any other dramatic biases in the feature data associated with human VISTA enhancers.

Next, we applied Step 2 of EnhancerFinder to all enhancer regions predicted in Step 1. We focused on brain, limb, and heart, because these tissues are highly represented in VISTA and have been extensively studied in previous analyses of developmental enhancers. We predicted 7,400 limb enhancers, 19,051 heart enhancers, and 11,693 brain enhancers (Figure 6) at a 5% FPR threshold tuned separately for each tissue. Since EnhancerFinder makes predictions for each tissue independently, there are no constraints on the distribution of tissues in the resulting genome-wide predictions. Nonetheless, we find a high level of tissue-specificity; 77% of the limb, heart, and brain enhancers are predicted to be active in just one of the three tissues.

All genome-wide enhancer predictions are available as tracks for import into the UCSC Genome Browser (Supplementary Data Files). These lists of tissue-specific enhancers should not be viewed as exhaustive; we found thousands of regions with positive, but less significant scores from Step 2 of EnhancerFinder.

**Predicted enhancers are associated with relevant functional genomic regions**
To characterize and further validate our genome-wide enhancer predictions, we examined their genomic distribution with respect to several independent indicators of function (details in Supplementary Text). Genes near brain and heart enhancers are enriched for expression in relevant tissues (Supplementary Tables 3 and 4). Similarly, Gene Ontology (GO) Biological Process enrichment analyses of the nearby genes suggest that our predicted developmental enhancers target genes that function in relevant cell types and tissues (Figure 6). The most prevalent transcription factor binding site motifs found in the sequences of predicted enhancers differed between enhancers of different tissues and included many relevant developmental TFs (Supplementary Table 5). Finally, our predicted enhancers contain 676 single nucleotide polymorphisms (SNPs) associated with significant effects in genome-wide association studies (GWASs) (Supplementary Table 6); this is significantly more overlap than expected at random (permutation $p < 0.001$).

Taken together, these analyses suggest that EnhancerFinder identifies many active regulatory regions that contain functionally relevant variation. Our tissue-specific enhancer predictions give valuable annotations to previously uncharacterized non-coding regions of the human genome. For example, thousands of SNPs associated with disease by GWASs are in non-coding regions with limited functional annotations [76]. Our genome-wide enhancer predictions provide a resource for exploring the mechanisms and functional effects of these uncharacterized GWAS hits.

***Case Study:* EnhancerFinder predictions highlight a novel enhancer near *ZEB2***
To illustrate how genome-wide enhancer predictions from EnhancerFinder can be used to facilitate the discovery of regulatory elements, we present a case study that identifies and validates a novel enhancer region near the human *ZEB2* gene, a zinc finger E-box-binding homeobox-2 TF (Figure 7). We selected the *ZEB2* locus for several reasons. *ZEB2* has many roles throughout embryonic and postnatal development, in particular in cortical neurogenesis [77,78,79,80], and mutations in *ZEB2* are associated with Mowat-Wilson syndrome, a complex developmental disorder [81]. However, relatively little is known about the genetic mechanisms that orchestrate *ZEB2*'s expression. A long-range enhancer of postnatal expression in developing kidney cells (E1 in Figure 7) was recently discovered 1.2 megabases (Mb) downstream of *ZEB2* in the adjacent gene desert [82]. Since this enhancer does not fully recapitulate the expression timing and domains of *ZEB2*, the authors speculate that the gene has many other, potentially long-range, enhancers. Supporting this theory, there are two validated E11.5 brain enhancers near *ZEB2* in the VISTA database (Figure 7, VISTA 407 and VISTA 1802). Finally, there is an enrichment of human accelerated regions (HARs) [83,84] near *ZEB2*, suggesting that it may have human-specific regulatory patterns.

Our enhancer predictions support the existence of a rich regulatory program specified in the non-coding sequence nearby *ZEB2*; there are 54 predicted enhancers for which it is the nearest TSS. This puts *ZEB2* in the top 0.2% of all genes with respect to the number of adjacent enhancer predictions. Supporting the validity of our predictions, the known VISTA enhancers both overlap EnhancerFinder predicted enhancers, while the regions tested by [82]and found to be not active or active at a later post-natal developmental stage (only E1) do not.

We selected the predicted enhancer indicated in the zoomed pane of Figure 7 for further experimental analysis due to its high EnhancerFinder score and overlap with a HAR. We interrogated the potential of the human and chimp sequences at this region to drive gene expression at E11.5 in transient transgenic mouse embryos. All seven embryos with staining showed cranial nerve expression (Figure 7 red box; Supplementary Figure 9), regardless of whether the construct contained the human or chimp sequence. Thus, we have identified a novel enhancer for *ZEB2*. Our enhancer predictions highlight many additional candidates for the further investigation of the regulation of *ZEB2*. We believe that our predictions will enable similar analyses of the regulatory landscape near many other genes of interest.

**DISCUSSION**

In this study, we developed EnhancerFinder, a new machine-learning framework for predicting regulatory enhancers from diverse data sources. In contrast to most previous enhancer identification strategies, which have based their predictions on one or a small number of data types, EnhancerFinder enables us to flexibly integrate the large and continually expanding collection of evolutionary, DNA sequence, and functional genomics data that are informative about enhancer function. Our analysis of the EnhancerFinder algorithm and its predictions makes three major contributions. First, we demonstrate that integrating diverse types of data from many cellular contexts, including some unexpected ones, improves prediction performance. Second, we show that enhancer tissue-specificity can be accurately predicted through a two-step approach in which tissues are individually evaluated after general enhancer prediction. Finally, our genome-wide developmental enhancer annotations, including tissue-specific predictions for heart, brain, and limb, highlight thousands of previously un-annotated genomic regions. We show that these predictions are enriched for a number of independent indicators of regulatory functions. As a result, we expect our predictions to prove useful in the annotation of non-coding genomic regions, as illustrated in the identification of a novel cranial nerve enhancer near *ZEB2*. The genome-wide predictions are freely available as a UCSC Genome Browser track.

*A biologically active in vivo definition of "enhancer."* We chose to define developmental enhancers for training as genomic regions that are experimentally shown to activate gene expression *in vivo* in embryonic mouse assays. This is a narrow definition, but we believe it is better suited to defining regions for further exploration and experimental characterization than approaches based on single data sources associated with enhancers. We show that our predicted enhancers, based on this biologically active definition, significantly overlap data sets commonly used as proxies for enhancer activity, such as H3K27ac and p300 binding. However, this does not imply that these other data alone are sufficient to identify enhancers, as we demonstrated for H3K27ac, H3K4me1, and p300 in Figure 3B. Similarly, when we evaluate the ability of previous computational methods for identifying enhancers, we find that they perform better than random, but significantly worse than EnhancerFinder, at identifying biologically active enhancers.

While EnhancerFinder can be used to predict enhancers in well-characterized cell lines, it is particularly useful at identifying enhancers in complex tissues that contain multiple cell types and in cell types that might not have much specific functional genomics data available. Other computational approaches to enhancer prediction have focused on identifying enhancers in individual cell types using functional genomics data from the same cells [57] or going one step further and using the differences in cell type specific transcription factor binding to identify cell-type specific binding motifs [62]. These methods generally perform well, but they do not address enhancer prediction in cell types with little or no functional genomics data, or in tissues that contain multiple cell types.

*Why do seemingly irrelevant data improve our enhancer predictions?* Data such as p300 binding sites and H3K4me1 have been used in previous studies to identify enhancers, and these data are major contributors to our enhancer predictions. However, data from many other sources and contexts less directly associated with enhancer activity provide complementary information that improves our predictions. Some of these data may be negatively correlated with enhancer activity, allowing EnhancerFinder to learn what features distinguish regions that are not developmental enhancers. Others may help reinforce patterns present in data from more relevant contexts, reflecting some degree of stability in the features of enhancer regions across developmental stages and cell types. For example, we found that features measured in embryonic stem cells are quite useful for E11.5 enhancer prediction; their removal from the classifier degrades performance and/or they have large (positive or negative) MKL weights. Examination of these features suggests that some identify "poised" regions that will become active enhancers upon differentiation, while others seem to help distinguish stem cell enhancers (i.e., non-enhancers at E11.5) from those specific to differentiated lineages. We note that despite these interesting observations, most individual features do not carry a great deal of information and the power of EnhancerFinder comes from the integration of many data sets. It is also possible that as a more complete experimental characterization of chromatin state and protein-DNA binding from E11.5 tissues is obtained, extra data from less relevant contexts will not provide as much improvement as it did in this study.

*What data are most informative about enhancer activity?* We focused on a single developmental stage with a large number of validated enhancers. To efficiently extend enhancer detection and validation to new contexts, it will be very important to select the most informative data to collect. Even though the ENCODE project has produced an impressive amount of data, it still has not extensively assayed most contexts of interest to researchers, in particular developmental biologists. The performance of classifiers trained on subsets of all our data and the weights we learned for feature sets and individual features provide some guidance for future experiments. Evolutionary conservation and DNA sequence patterns are broadly useful in the identification of enhancers, but our results suggest that adding functional genomics data is necessary to make more precise predictions about the contexts of activity. H3K4me1 and p300 are two of the most useful functional genomics data types overall (Supplementary Figure 5), but many others are useful in particular contexts.

*Why are heart enhancers easier to predict than other types of enhancers?* We saw a broad range in our ability to predict the tissue specificity of enhancers from existing data. Heart enhancers were dramatically easier to identify than other tissue-specific enhancers. Heart enhancers have significantly higher GC content than enhancers of other tissues, are less evolutionarily conserved, and are closer to the nearest TSS than other known enhancers at E11.5, and we show that GC content alone is sufficient to accurately predict many heart enhancers (Supplementary Figures 6 and 7). The underlying biological explanation for these patterns may have to do with relative developmental age of different organs and tissues. At E11.5, the heart is further along its developmental trajectory than the other tissues considered. At this stage, many of the less conserved, mammal-specific features of the heart are developing [85,86], whereas other tissues are still developing under more general, less species-specific conserved regulatory programs at E11.5 [87]. A recent study of enhancers in the adult mouse retina found that high local

GC content was strongly correlated with enhancer activity [88]. Paired with our result, this suggests that GC content is a distinguishing feature of certain classes of enhancers.

*Limitations of our approach.* In spite of the strong overall performance of EnhancerFinder at predicting tissue-specific developmental enhancers, our approach has some limitations. First, we rely heavily upon the VISTA Enhancer Browser for training examples, because it is the largest collection of validated mammalian enhancers currently available. This resource provides an impressive catalog of validated human regulatory enhancers, but it is limited to a single developmental stage and experimental system. Without more data and analysis, it is difficult to evaluate how specific our predictions are to this context. Applying EnhancerFinder to known enhancers in model organisms, such as zebrafish and fly, would provide additional opportunities to evaluate our approach and findings, while potentially demonstrating differences in how enhancers function in these different species.

Second, most of the enhancers present in VISTA are evolutionarily conserved. As a result, the VISTA enhancers cannot be viewed as an exhaustive catalog of the full range of enhancers. However, these regions have validated enhancer activity *in vivo*, and thus provide an appealing alternative to approaches that use single-mark proxies for enhancer activity (e.g., considering all H3K27ac peaks as active enhancer regions). We mitigated the impact of the bias towards conserved regions by removing evolutionary conservation as a feature from EnhancerFinder when we applied it to predict enhancers genome-wide. Our goal in doing so was to improve our ability to discern less conserved enhancers in these genome-wide predictions. This enabled the identification of thousands of non-conserved enhancers (~20% of all predictions).

Third, though our predictions are based on a large collection of genome-wide chromatin state, protein-binding, and sequence information from many contexts, we are still limited by data availability. Even with the impressive efforts of ENCODE and related projects, producing data that are perfectly matched to all contexts of interest is time consuming and sometimes impossible, especially when studying humans. Thus, it will be important to develop a principled understanding of how different data can be generalized across tissues, developmental stages, and between species. In our analysis, the highest weighted features come from contexts close to the developmental stage of interest, and thus we anticipate that gathering more data from developmentally relevant cells and tissues will significantly improve our ability to annotate genomic regions involved in the regulation of embryonic development. However, data from other, seemingly unrelated, contexts may continue to prove useful.

*Extensions and future applications.* This study makes a significant contribution towards annotating regulatory elements in the human genome and provides tools for interpreting the effects of mutations in non-coding regions. Our case study on the region around *ZEB2* illustrates how our predictions can facilitate the rapid identification of novel enhancers. In addition, the statistical enrichment for GWAS SNPs in our genome-wide enhancer predictions suggests that they may be a good resource for pinpointing causal mutations in potential disease loci.

EnhancerFinder is a general framework for enhancer prediction and rigorous evaluation of different data sources that aim to annotate the regulatory functions of the human genome. It could easily be extended to include additional types of data, such as population-level variation at each locus, information about the three-dimensional state of the genome from Hi-C and 5C, and predictions of potential target genes for each enhancer. It could also be used to analyze additional aspects of the data we already consider, such as accounting for the relative genomic position of different features [67].

The EnhancerFinder two-step approach enables delineation of features common to all enhancers versus those that characterize enhancers of different types. For example, we find that predicting the tissue specificity of enhancers that are unique to a single tissue is more difficult than those that are active in multiple tissues (Supplementary Figure 8), that certain features make prediction of heart enhancers particularly easy, and that different features are selected in classifiers for general enhancers and those for specific tissues. Together, these results suggest that there may be distinct classes of enhancers, even amongst those active in a given tissue at a single developmental stage. Further analysis of EnhancerFinder classifiers based on different types of data may help suggest biological mechanisms underlying the functional distinctions and genomic features of these different classes of enhancers.

**METHODS**

All work presented in this paper is based on the February 2009 assembly of the human genome (GRCh37/hg19) downloaded from the UCSC Genome Browser (http://genome.ucsc.edu/). Any data that was not in reference to this build was mapped over using the *liftOver* tool from the UCSC Kent tools (http://hgdownload.cse.ucsc.edu/admin/jksrc.zip).

**Multiple kernel learning-based prediction of developmental enhancers**
In our framework, genomic regions are associated with a common set of descriptive features. We then apply machine-learning algorithms that use the features of known training examples to learn a function of the feature data that distinguishes the positives (enhancers) from the negatives (non-enhancers). This function can then be applied to the features associated with uncharacterized genomic regions to predict their enhancer status. A positive score for a genomic region indicates predicted membership in the positive class (enhancers) and a negative score indicates predicted membership in the negative class (non-enhancers).

*Training examples*
We obtained all of our positive training data and our tissue-specific negative training data from the VISTA Enhancer Browser [70] on April 4, 2012. We downloaded the location, DNA sequence, and expression contexts for all human sequences tested in the VISTA mouse E11.5 enhancer screen. This consisted of 711 validated human enhancers and 736 genomic regions that did not exhibit enhancer activity in this context (http://enhancer.lbl.gov/). The median length of the enhancers in VISTA is 1,545 bp.

In the first step of EnhancerFinder (Figure 1), we used all 711 VISTA enhancers as positive training data. For negative training data, we generated a set of 711 random genomic regions matched to the length and chromosome distribution of the positives, and filtered to remove known VISTA enhancers and assembly gaps.

In the second step of EnhancerFinder, we used tissue-specific subsets of the 1,447 VISTA regions for positive and negative training data. For example, when predicting heart enhancers, our positive training data were the 84 VISTA regions with heart expression in E11.5 mice, and our negative training data were the remaining 1,363 VISTA regions that were tested and showed no heart expression at E11.5, even though they may be enhancers in other tissues. We did not require that a region be active only in the tissue of interest. We included the VISTA negatives in this analysis, because they share many attributes in common with known enhancers and may have enhancer activity in contexts other than E11.5. Our results did not change dramatically when the VISTA negatives were not included in the training. We trained tissue-specific classifiers for the six tissues with more than 50 examples in VISTA: forebrain, midbrain, hindbrain, heart, limb, and neural tube. We also trained a brain enhancer classifier on the combined the forebrain, midbrain, and hindbrain enhancers.

*Feature data*
We consider three main types of data as features in our analysis: functional genomics data, evolutionary conservation, and DNA sequence motifs. We obtained our functional genomics feature data from the ENCODE data repository at the UCSC Genome Browser (http://genome.ucsc.edu/ENCODE/ and [89]). These data include histone modifications, such as H3K4me1, H3K4me3, H3K27ac, protein-DNA associations for many TFs and p300, and several measurements of open chromatin (DNaseI hypersensitivity, FAIRE, digital genomic footprinting), from hundreds of cell types [89]. We also included heart p300 data from [40]. For a full list of the functional genomics data considered, see Supplementary Table 1. We associated each genomic region with a binary vector that represents the presence or absence of overlap with each functional genomics data set. To determine this feature vector, we intersected the genomic location of the region of interest with the peaks defined by the original researchers (from the broadPeak or narrowPeak files) using *intersectBed* [90]. We found that considering non-binary functional genomics features based on experimental data, like the density of sequence reads from a ChIP-seq study, did not significantly improve performance (data not shown). However, we suspect that with consistent peak calling and appropriate normalization this might be an avenue for future improvement.

To summarize the DNA sequence motif patterns in a genomic region, we calculated the number of occurrences of all possible 4-mers in the sequence.

Evolutionary conservation estimates were taken from the mammalian phastCons elements [73] obtained from the *phastConsElements46wayPlacental* track in UCSC Genome Browser. Each genomic region was assigned its maximum overlapping phastCons score or zero if it did not overlap any phastCons elements.

*Machine-learning algorithms*

EnhancerFinder is an extension of the SVM supervised learning framework that allows the integration of multiple data types into a single discrimination function. Standard 1-norm MKL augments the usual SVM discrimination function, $f$, with additional parameters, $\beta_j$, that weight the contribution of each kernel function $k_j$:

$$f(\mathbf{x}) = \sum_{i=1}^{N} \alpha_i \sum_{j=1}^{M} \beta_j k_j(\mathbf{x}, \mathbf{x_i}) + b$$

where $N$ is the number of training examples, $M$ is the number of kernels, $\alpha_i$ are the training example weights, and $b$ is the bias [67]. We include three kernel functions in EnhancerFinder, each of which corresponds to one of the three types of feature data described above. These kernels quantify the similarity of the features of the appropriate type for any two genomic regions. To combine the kernels, the MKL algorithm simultaneously learns weights for the associated kernels, in addition to learning the bias and weights for each training example as in a standard SVM. We use the 4-spectrum kernel [72] for our sequence features; this kernel has been shown to perform well in a variety of DNA sequence-based prediction tasks including enhancer prediction [55]. For the functional genomics and evolutionary conservation data, we use linear kernels, which are equivalent to dot products of the feature vectors. We investigated using more sophisticated kernel functions for these features, but Gaussian and polynomial kernels did not yield any consistent improvement. Each kernel was variance normalized, and we balanced the misclassification costs by class size [91]. In addition to EnhancerFinder classifiers, we also trained and evaluated the constituent single kernel SVMs. All analyses were performed using the implementation of SVMs and MKL in the SHOGUN Machine Learning Toolbox v1.1.0 [92].

**Performance evaluations**

To evaluate the performance of trained classifiers, we performed 10-fold cross-validation on the training data and quantified our results with ROC AUC, precision-recall curves, and power estimates at fixed false positive rates. We computed p-values for the difference in performance between classification methods using McNemar's test [93,94]. To estimate false discovery rates, we trained EnhancerFinder classifiers at 1:1, 1:10, and 1:100 ratios of positive to negative enhancers and used the resulting 10-fold cross-validation results to calculate the proportion of false discoveries genome-wide at a 5% FPR if the true proportion of 1.5 kb windows containing an enhancer was 50%, 10%, or 1%.

**Comparison to existing enhancer prediction methods**

We compared EnhancerFinder's predictions to those of several previous enhancer prediction methods. We obtained the performance of CLARE on our Step 1 prediction task, by inputting our positive and negative data into the CLARE web server [74]. We downloaded the genomic segmentations and annotations produced by ChromHMM [65] and Segway [66]. We considered the ChromHMM predictions based on different ENCODE cell lines both individually and together. Any genomic region in our evaluation data set that overlapped an enhancer state was considered a predicted enhancer, and all others were considered predicted non-enhancers. For Segway, we also considered the "TF activity" state.

**Identification of tissue-specific enhancers across the human genome**

We predicted tissue-specific developmental enhancers throughout the human genome by applying a trained MKL classifier (Step 1 of EnhancerFinder) without conservation (see Results) to sliding windows of 1500 bp, moving along the human genome in 500 bp steps. The feature profile for each window was computed as described above. To focus on high-confidence predictions, we filtered the enhancer scores for the windows at a 5% FPR, estimated from cross-validation using the genomic background, and combined the remaining overlapping windows to produce 84,301 high-confidence predicted enhancers.

To predict tissue specificity, we applied trained brain, limb, and heart classifiers (Step 2 of EnhancerFinder) without conservation to all 299,039 windows with positive enhancer scores in Step 1. We then applied a 5% FPR cutoff for each tissue and concatenated the remaining overlapping windows into merged enhancer regions. Using this approach, we predicted 19,051 heart enhancers, 11,693 brain enhancers, and 7,400 limb enhancers.

**Analysis of genome-wide tissue-specific enhancer predictions**

We characterized the expression patterns of the gene nearest to each predicted enhancer using the GNF Atlas 2 [95]. It contains expression data for genes in 79 different tissues, with expression measured using Affymetrix microarrays. For each of these 79 tissues, we used a paired t-test to determine if the nearest genes of predicted heart enhancers had significantly different mean values of expression than the nearest genes of brain enhancers. We did not include the limb enhancers in this analysis due to the lack of relevant expression data in the GNF Atlas 2.

We examined genomic regions near predicted developmental enhancers for enrichment of Gene Ontology functional annotations, known phenotypes, and pathways using GREAT [96]. Results were computed using the hypergeometric test for genome-wide significance, with the default settings and the "basal plus extension" association rule (proximal 5 kb upstream, 1 kb downstream, plus distal up to 100 kb).

We identified the sequence motifs present in each set of enhancers using the FIMO tool (Find Individual Motif Occurrences) from the MEME Suite of sequence motif analysis tools [97]. We considered known transcription factor binding motifs from the April, 2011 release of the TRANSFAC database with a FIMO score threshold of 10e-5. We identified those occurrences that fell in predicted enhancers, and summarized motifs to identify the most prevalent TFs in each tissue-specific set of enhancers.

We analyzed the overlap of predicted enhancers with GWAS SNPs, based on the NHGRI catalog of 9,687 GWAS SNPs downloaded from the UCSC Genome Browser in October 2012. Unadjusted permutation p-values were calculated by randomizing genomic locations of predicted enhancers (matching for length and chromosome, and avoiding assembly gaps) and overlapping these randomized regions with GWAS SNPs to assess significance of overlapping regions.

**Transgenic Enhancer Assays**

Enhancer assays were carried out in transient transgenic mouse embryos generated by pronuclear injections of expression constructs into FVB embryos (Cyagen Biosciences). Human and chimpanzee DNA sequences were inserted upstream of a minimal promoter and a *LacZ* reporter gene. The human sequence was amplified using primers 5'-TGTATGAAACCTGTTCACTCTCC-3' and 5'-GCTTAAAACAACTACTAGAATCAGGC-3' from the bacterial artificial chromosome (BAC) RP11-107E5 (from the BacPac resource at CHORI). The chimpanzee sequence was amplified using primers 5'-TGTATGAAACCTGTTCACTCTCC-3' and 5'-GCTTAAAACAACTACTAGAATCAGGC-3' from BAC CH251-677E03a (CHORI). The embryos were collected and stained for *LacZ* expression at E11.5.

Following the annotation policies of the VISTA Enhancer Browser, we required that consistent spatial expression patterns be present in three or more embryos with staining in order for the region to be considered an enhancer.

**ACKNOWLEDGMENTS**

We thank A. Robles and C. Miller in the Gladstone Histology Core for assistance with embryo imaging and J. Rubenstein for help interpreting embryo staining patterns. Transgenic mice were generated by Cyagen Biosciences, Inc. This project was supported by grants from NIGMS (#GM082901) and NHLBI (#HL098179), a PhRMA Foundation fellowship, a University of California Achievement Awards for College Scientists (ARCS) Scholarship, a gift from the San Simeon Fund, and institutional funds from the J. David Gladstone Institutes.

**FIGURE CAPTIONS**

**Figure 1. Overview of the EnhancerFinder enhancer prediction pipeline.** In our two-step approach, regions of the genome are characterized by diverse features, such as their evolutionary conservation, regulatory protein binding, chromatin modifications, and DNA sequence patterns. For each step, appropriate positive (green) and negative (purple) training examples are provided as input to a multiple kernel learning (MKL) algorithm that produces a trained classifier (black box). We used 10-fold cross validation to evaluate the performance of all classifiers. In Step 1, we trained a classifier to distinguish between known developmental enhancers from VISTA and the genomic background. In Step 2, we trained several classifiers to distinguish enhancers active in tissues of interest from those without activity in the tissue according to VISTA. We applied the trained enhancer classifier from Step 1 to the entire human genome to produce more than 80,000 developmental enhancer predictions. We then applied the tissue-specific enhancer classifiers from Step 2 to further refine our predictions.

**Figure 2. Combining diverse data using EnhancerFinder improves the identification of enhancers.** (A) Enhancer prediction strategies based on functional genomics data, evolutionary conservation, and DNA sequence motif patterns all perform well, but EnhancerFinder, which combines these data, provides significant improvement over each of them alone (p<2.0E-7 for all). (B) Each of the approaches from (A) predicts that somewhat different sets of the VISTA regions are enhancers. This suggests that complementary information is contained in each data source. EnhancerFinder (not shown), which combines them, captures many of the enhancers that are unique to each source; it predicts 25 of the 44 enhancers unique to **Functional Genomics**, 30 of the 76 unique to **DNA Sequence Motifs**, and 34 of the 111 unique to **Evolutionary Conservation**. (C) EnhancerFinder also outperforms several other published approaches to enhancer prediction. CLARE is a machine learning method based on known regulatory motifs. ChromHMM and Segway are unsupervised clustering methods that have been used to segment the genome into different functional states based on patterns in functional genomics data. The enhancer predictions of ChromHMM and Segway do not distinguish VISTA enhancers from the genomic background as well as our machine learning approaches. Each "X" gives the performance achieved at the Step 1 enhancer classification problem obtained by considering any region that overlaps a ChromHMM enhancer state a positive and all others a negative. The different X's represent state predictions based on data from different ENCODE cell types: GM12878 (blue), H1-hESC (violet), HepG2 (brown), HMEC (tan), HSMM (gray), HUVEC (light green), K562 (green), NHEK (orange), NHLF (light blue), and all contexts combined (red). The circles give the performance of the Segway "enhancer" and "TF activity" states.

**Figure 3. Integrating diverse functional genomics data improves enhancer prediction.** (A) Considering all functional genomics features, including those from contexts and assays not expected to be associated with developmental enhancer activity (**All Functional Genomics** and **Relevant Functional Genomics**), improves the ability of the classifiers to identify developmental enhancers (**Embryonic Functional Genomics**; p=9.2E-9 and p=2.7E-6, respectively). (B) Combining H3K4me1, p300, and H3K27ac

data, which are commonly used in isolation to identify enhancers, in a linear SVM (**Basic Functional Genomics**) is better able to distinguish known developmental enhancers from the genomic background than considering each type of data alone (p<2E-7, for each). However, combining these marks still performs significantly worse than EnhancerFinder (Figure 2A; AUC=0.96).

**Figure 4. Enhancers of heart expression are easier to identify than enhancers active in other tissues at E11.5.** (A) We trained EnhancerFinder using the same features as in Step 1 (Figure 1), but using VISTA enhancers active in a given tissue as positives and tested regions that did not show activity in the tissue as negatives (Step 2). Heart enhancers were dramatically easier to identify than enhancers of expression in other tissues. The heart enhancers have significantly higher GC content than other enhancers and the genomic background, and this largely explains the ease of identifying them (Supplementary Figure 7). In general, EnhancerFinder improves on it component methods when predicting tissues of activity, but functional genomics data alone often perform competitively (Supplementary Table 2).

**Figure 5. EnhancerFinder's two-step approach captures tissue-specific attributes of enhancers.** (A) The true overlap of human enhancers of brain, heart, and limb in the VISTA database. The vast majority of enhancers are unique to one of these tissues at this stage. For example, of the 84 validated heart enhancers, 71 are unique to heart, five are shared with brain, four with limb, and four with both. (B) The predicted overlap of VISTA enhancers based on predictions made with a single training step using MKL with only enhancers of that tissue considered positives and the genomic background as negatives. This approach overestimates the number of enhancers active in multiple tissues. Each classifier mainly learns general attributes of enhancers, rather than tissue-specific attributes. (C) The predicted overlap based on EnhancerFinder's two-step approach. These predictions are much more tissue-specific and exhibit overlaps between tissues similar to the true values (A). Tissue distributions for the classifiers in (B) and (C) are similar when they are applied to other genomic regions, as illustrated in our genome-wide predictions; only predictions overlapping VISTA enhancers are shown here to enable comparisons to the distribution for validated enhancers (A).

**Figure 6. Predicted tissue-specific enhancers exhibit tissue-specific characteristics.** EnhancerFinder identifies thousands of novel high-confidence (FPR < 0.05) heart, brain, and limb enhancers**.** These enhancers are enriched for tissue-specific GO Biological Processes. The five most enriched GO Biological Processes among genes near each enhancer set (as calculated using GREAT) are listed in the colored boxes. Over 70% of EnhancerFinder predicted heart, brain, and limb enhancers are unique to a single tissue. The larger number of high-confidence heart enhancers relative to brain and limb enhancers is the result of the superior performance of the heart classifier.

**Figure 7. A novel cranial nerve enhancer in the *ZEB2* locus.** This screen shot from the UCSC Genome Browser shows a dense region of predicted enhancers in a 1.5 Mb window on human chromosome 2 including *ZEB2* and part of the adjacent gene desert. Tracks give the locations of four human accelerated regions (HARs), two validated

VISTA enhancers, and the E1 region recently shown to have postnatal enhancer activity by [82]. The inset shows a zoomed in view of *ZEB2* (hg19.chr2:145,100,000-145,425,000) along with summaries of several ENCODE functional genomics datasets and evolutionary conservation across placental mammals. We tested the predicted enhancer overlapping 2xHAR.240 for enhancer activity at E11.5 in transgenic mice. Both the human and chimp versions of this sequence drive consistent expression in the cranial nerve (Supplementary Figure 9).

## SUPPLEMENTARY FIGURE CAPTIONS

**Supplementary Figure 1. Precision-Recall curves corresponding to all ROC curves presented in the main text.** (A) Figure 2A (B) Figure 2C The CLARE method, which is included in main text Figure 2C, was not included in this corresponding figure because we could not obtain the raw scores from regions from the web server (C) Figure 3A (D) Figure 3B (E) Figure 4.

**Supplementary Figure 2. The 4-spectrum kernel performs competitively with other k-spectrum kernels and the combination of k-spectrum kernels.** We analyzed the ability of spectrum kernels based on k-mer lengths between 2 and 8 to distinguish enhancers from the genomic background (Step 1). K-mers between 4 and 7 had the best performance. We also evaluated an MKL algorithm that combined each k-spectrum kernel, and it did not provide significant improvement over the best individual kernels.

**Supplementary Figure 3. Considering known TFBS motifs does not improve the 4-spectrum kernel.** Considering the number of occurrences of known TFBS motifs as features has recently been used in a linear SVM framework to predict enhancers [53]. To evaluate the utility of this approach, instead of and in addition to considering all k-mers, we created a linear SVM that used the number of hits to 1022 TF binding site matrices from TRANSFAC and JASPAR as computed by FIMO as features. That is the feature vector for each region consisted of 1022 elements, each of which was the number of significant hits for a different TF motif. This TFBS linear SVM (AUC=0.81) did not perform as well as the 4-spectrum kernel (AUC=0.88). We also evaluated an MKL algorithm that combined the 4-spectrum and TFBS kernels. This combined kernel did not perform any better than the 4-spectrum kernel suggesting that, at least under this encoding, TFBS motifs do not provide significant additional benefit in distinguishing enhancers from the genomic background.

**Supplementary Figure 4. Combining functional genomics data with an SVM outperforms simply considering regions overlapping these data.** The four solid lines shown are the same as in Figure 4B; they summarize the performance of these methods at distinguishing VISTA enhancers from the genomic background (Step 1). The X's give the performance of an approach that considers all regions overlapping a given feature as positives and all others as negatives. The + and * indicate the performance obtained by considering the union and intersection of H3K4me1, p300, and H3K27ac, respectively.

For each feature, the linear SVM achieves better performance than simply considering all overlapping regions as positives.

**Supplementary Figure 5. EnhancerFinder feature weights highlight the contribution of different functional genomics data types to enhancer predictions.** Each "+" represents the contribution made by a single data feature, e.g. H3K4me1 peaks from embryonic stem cells, to the classification in EnhancerFinder Step 1 (developmental enhancers versus genomic background). Positive weights (red) indicate an association with enhancer activity in our analysis and negative weights (blue) suggest a lack of enhancer activity. The features plotted here come from a range of likely relevant contexts (**Relevant Functional Genomics** classifier; Supplementary Table 1), and the number of data sets present for each feature type is given in parentheses. The black bar gives the average weight over all features of each type. In general, the features with high average weights, such as H3K3me1, p300, and H3K4me2, are known to be associated with enhancers, while those with large negative weights are associated with other types of genomic regions. However, no data type has uniformly positive or negative weights in all contexts.

**Supplementary Figure 6. Heart enhancers are less conserved and closer to the nearest transcription start site (TSS) than limb and brain enhancers.** Considering only limb and brain enhancers that are less evolutionarily conserved and close to a TSS improved our ability to identify them, but they are still more difficult to identify than heart enhancers. The high GC content of heart enhancers proved essential to the ease of predicting them (Supplementary Figure 7).

**Supplementary Figure 7. The uniquely high GC content of heart enhancers in VISTA enables accurate classification.** The VISTA heart enhancers have higher GC content (49%) than other types of enhancers and the genomic background (~40%). (A) The classification score from a spectrum kernel classifier trained to distinguish heart enhancers within VISTA (Step 2) is strongly correlated (Pearson rho=0.95) with the GC content of the input region. (B) A classification algorithm based solely on GC content (black) performs competitively with the spectrum kernel (AUC of 0.80 vs. 0.82), and nearly as well as EnhancerFinder (0.85; Figure 5).

**Supplementary Figure 8. Enhancers active in multiple tissues are easier to identify than those active in a single tissue.** There are 399 enhancers active in a single tissue at E11.5 in the VISTA database and 312 active in multiple tissues. EnhancerFinder is better able to distinguish the enhancers active in multiple tissues from the VISTA negatives (AUC=0.75) than it is to distinguish single tissue enhancers from the negatives (AUC=0.67). This trend also holds across each tissue individually. However, both sets are easy to distinguish from the genomic background (AUC=0.96 for both, not shown).

**Supplementary Figure 9. Transient transgenic mouse embryos support a novel cranial nerve enhancer near *ZEB2*.** Seven transient transgenic mouse embryos showed *LacZ* expression at embryonic day 11.5. Constructs containing a 999 bp region (hg19.chr2:145,234,541-145,235,539) including 2xHAR.240 near *ZEB2*, a minimal

promoter, and *LacZ* were used for human. The orthologous region was used in the chimp construct (panTro2.chr2b:148,811,929-148,812,929). Three embryos with constructs containing the human version of the region of interest and four embryos containing the chimp sequence had staining. In all embryos, there is consistent expression in the cranial nerve. There does not appear to be a significant difference between human and chimp at this time point.

**MKL training:**

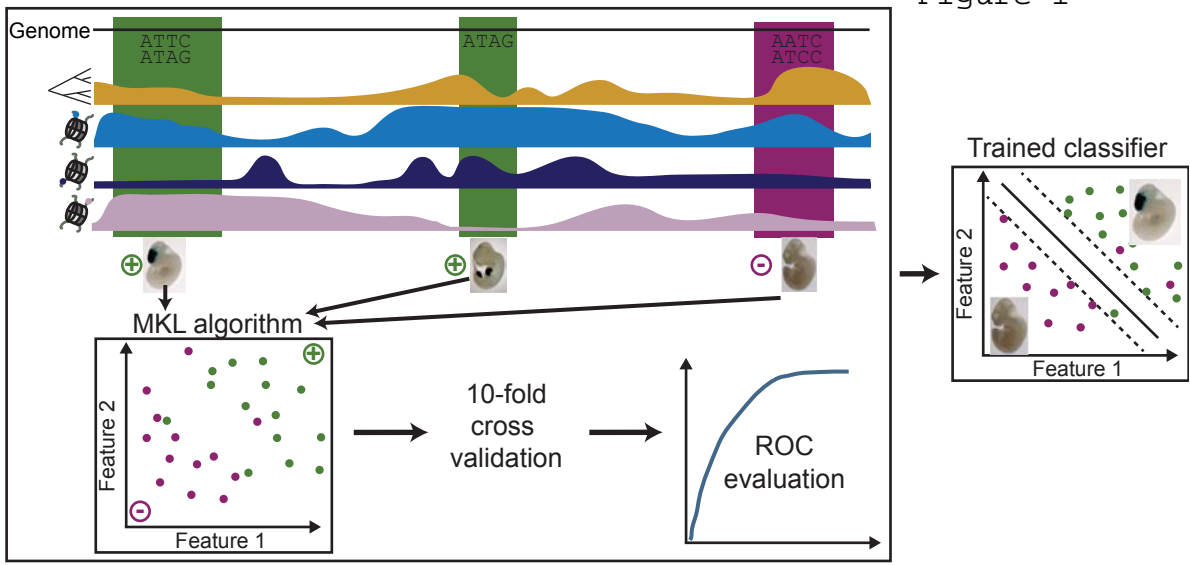

**Genome-wide enhancer prediction:**

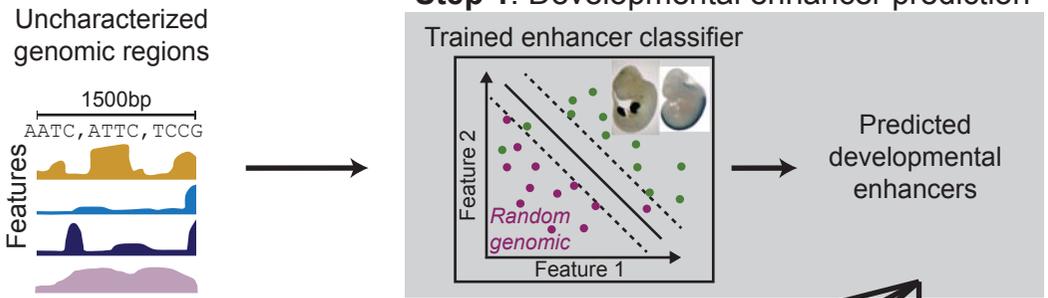

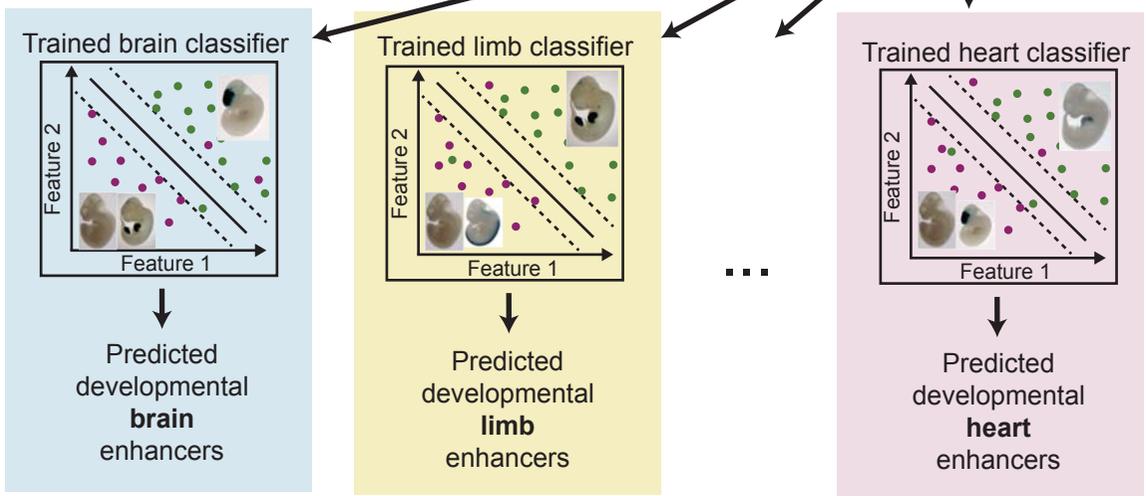

**A**

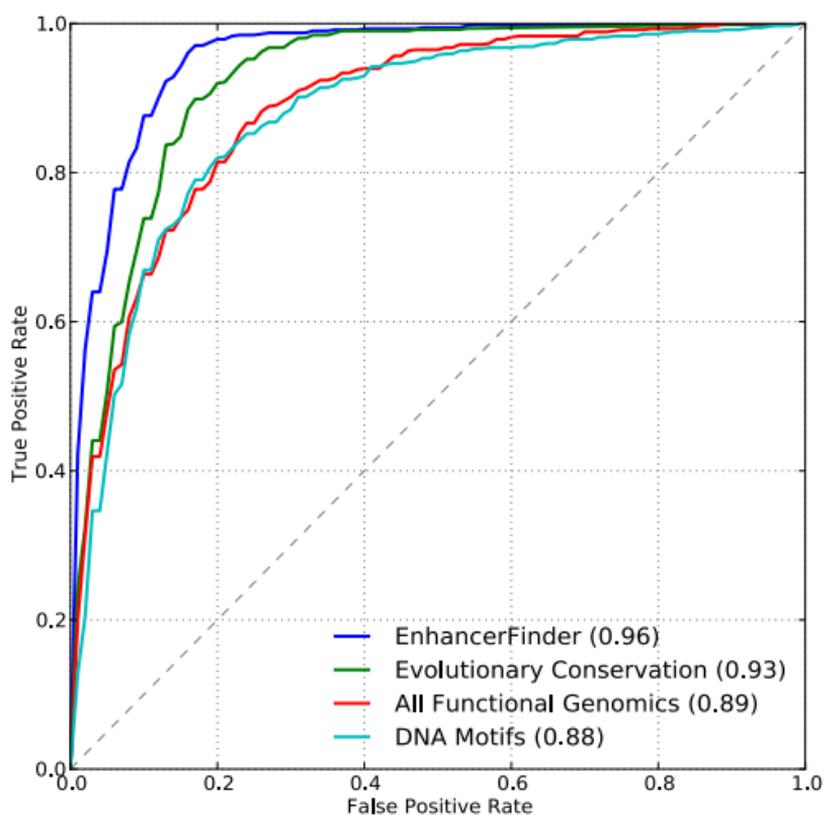

**B**

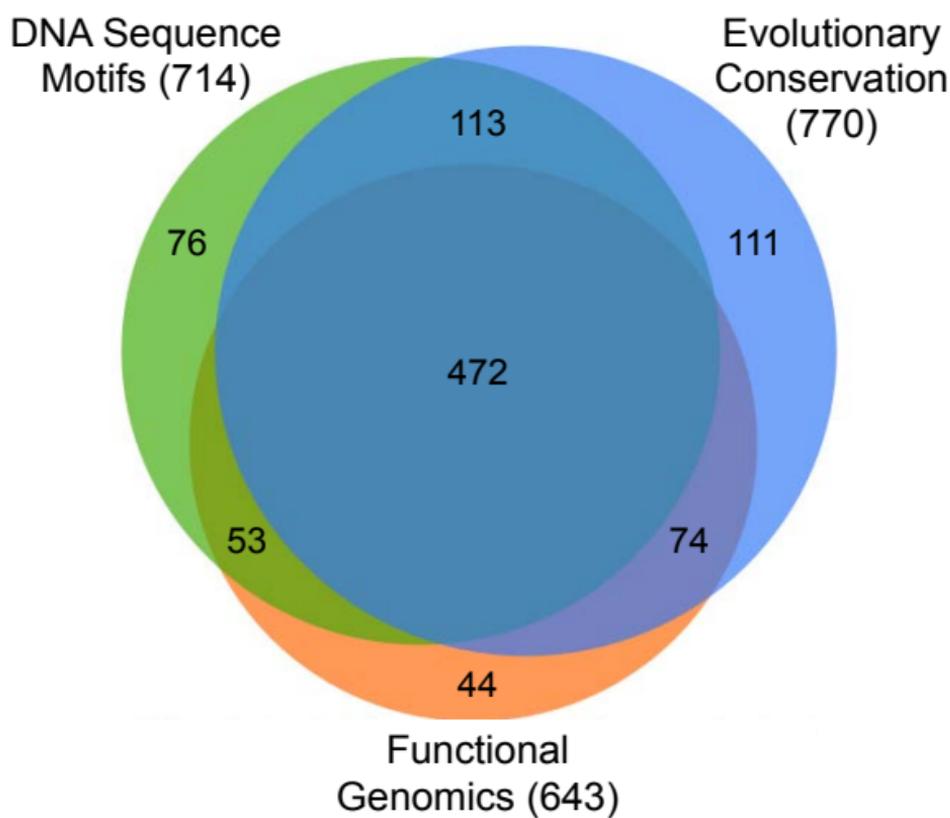

**C**

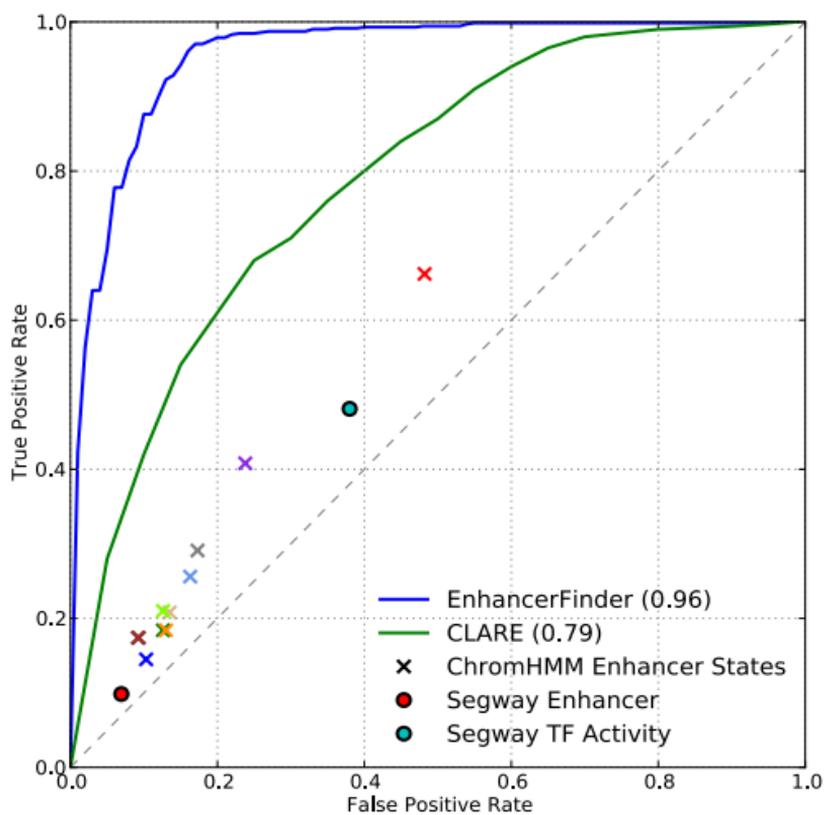

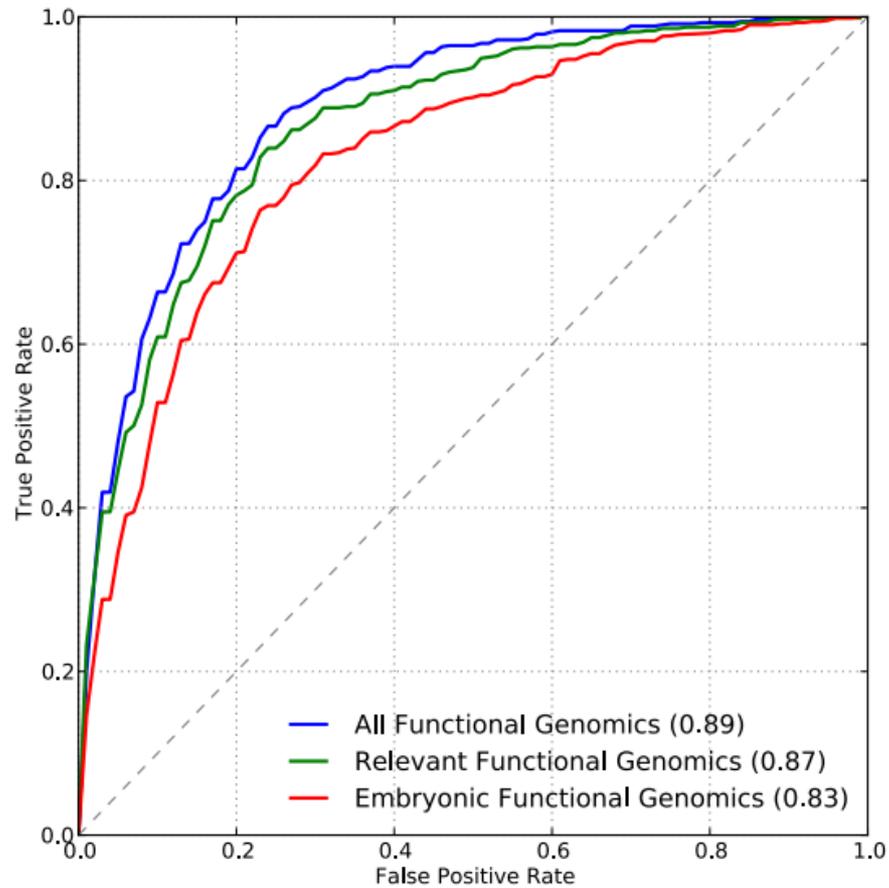

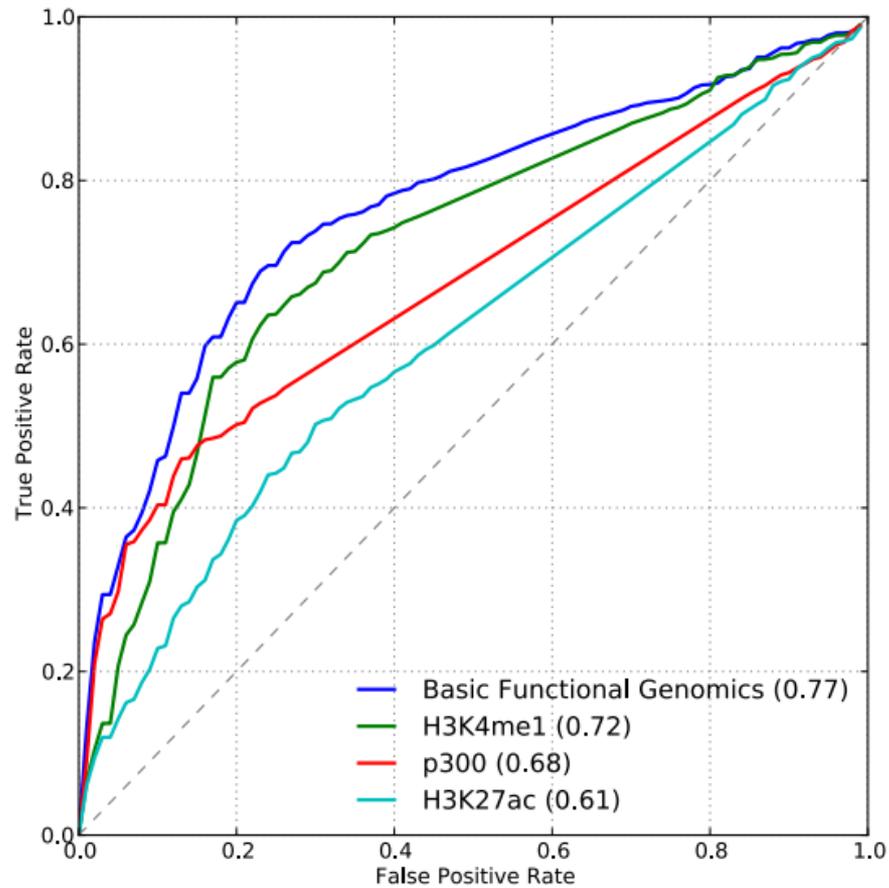

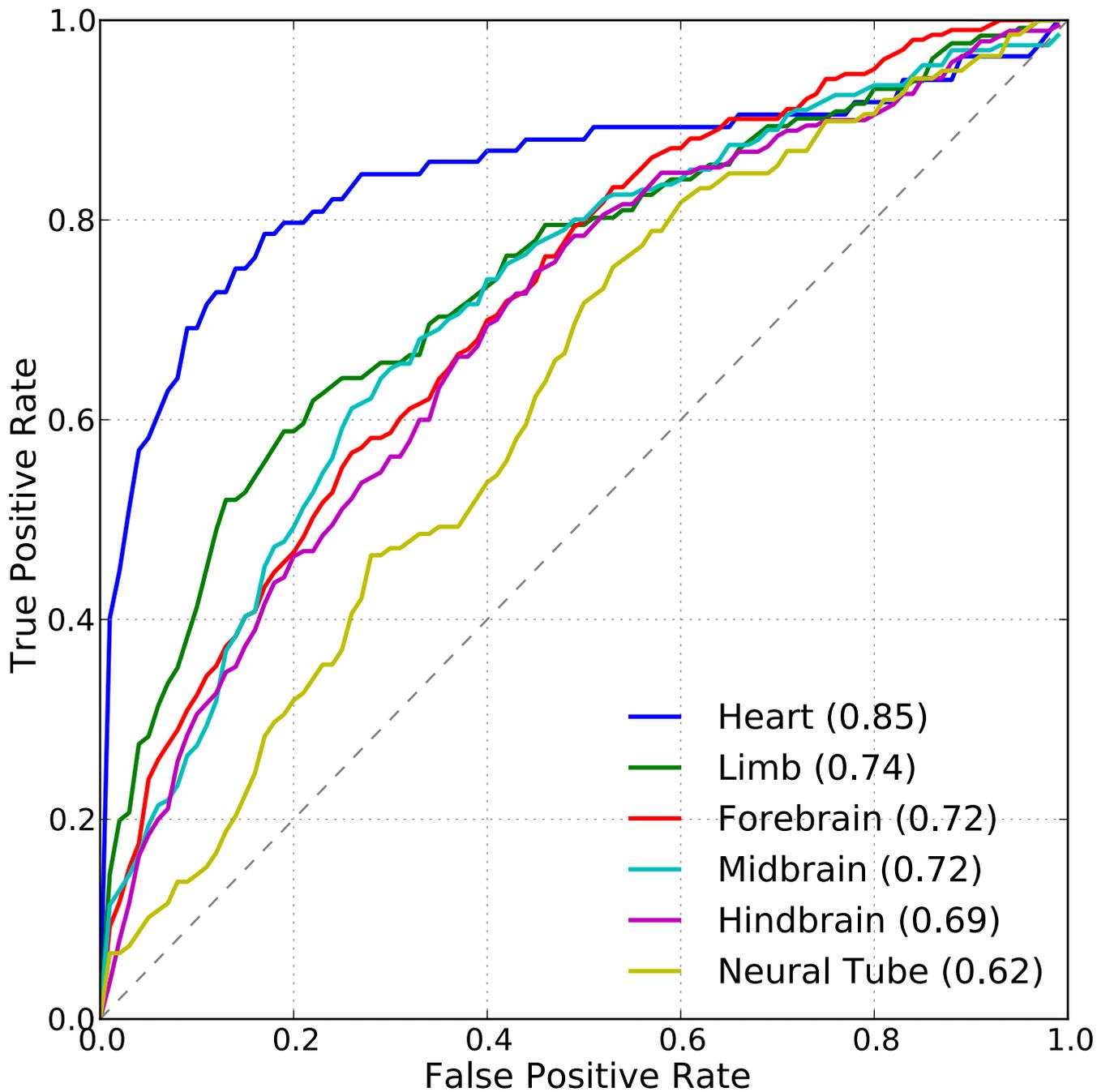

**A** **VISTA Positive Tissue Overlap**

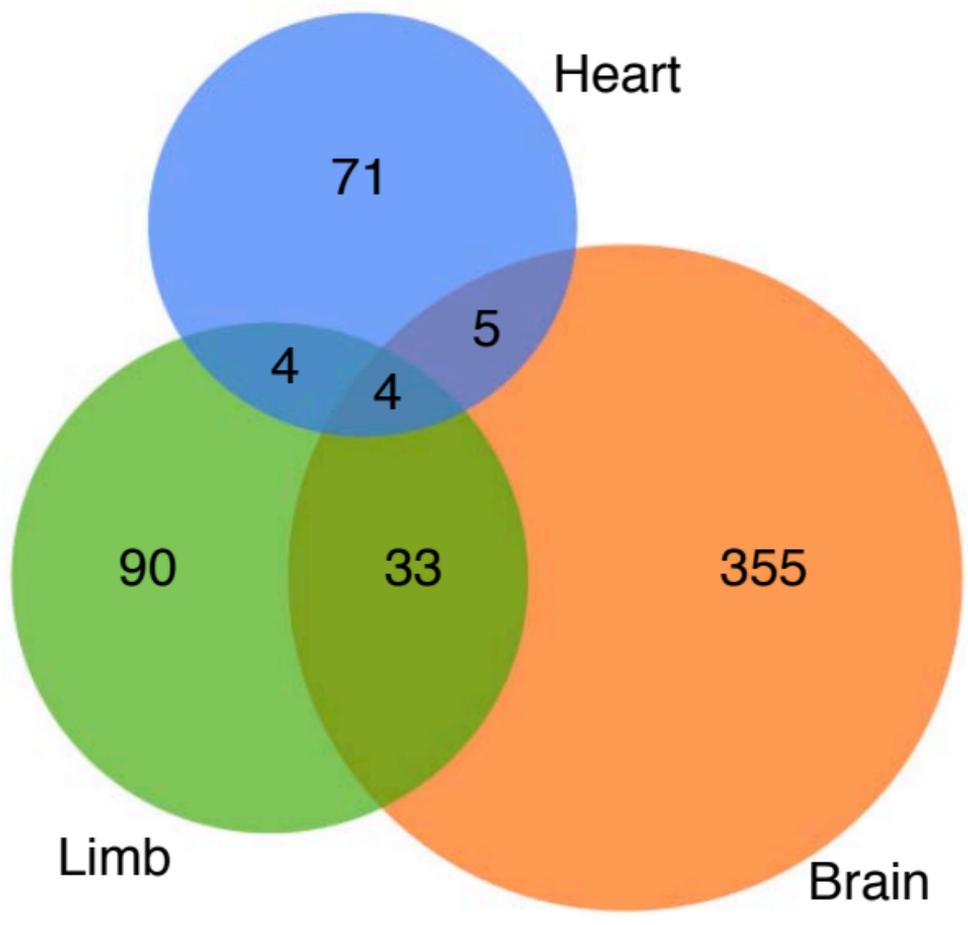

**B** **One Step Prediction Tissue Overlap**

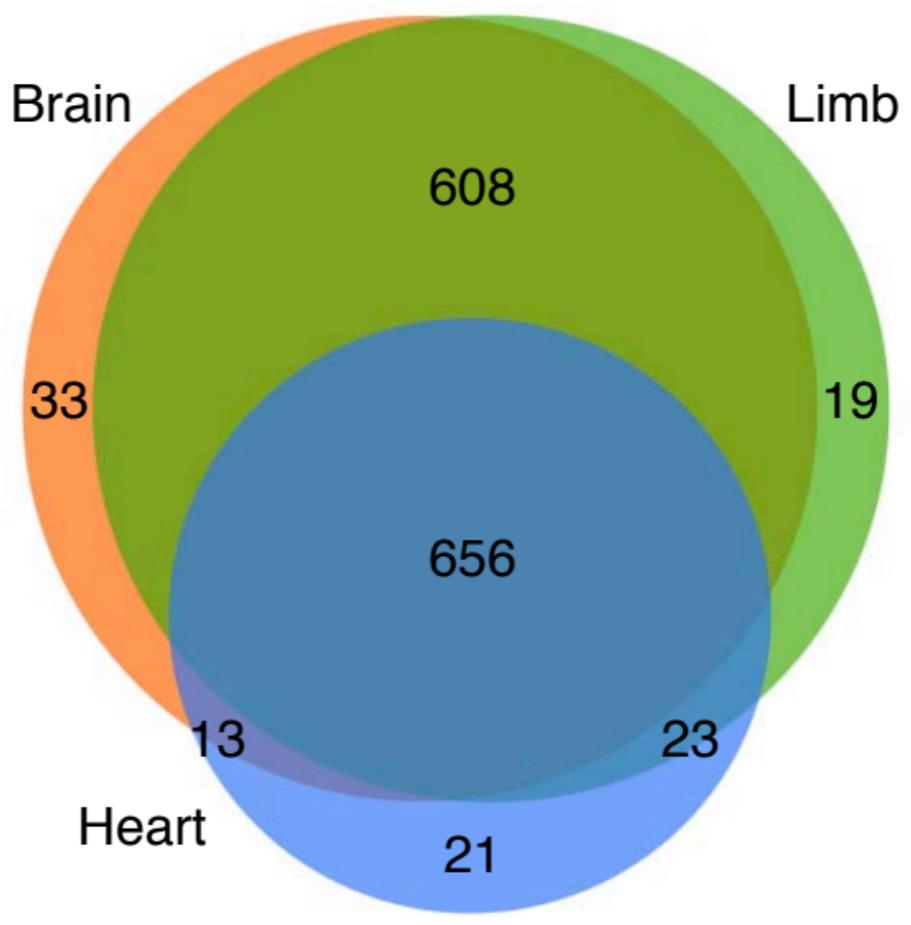

**C** **Two Step Prediction Tissue Overlap**

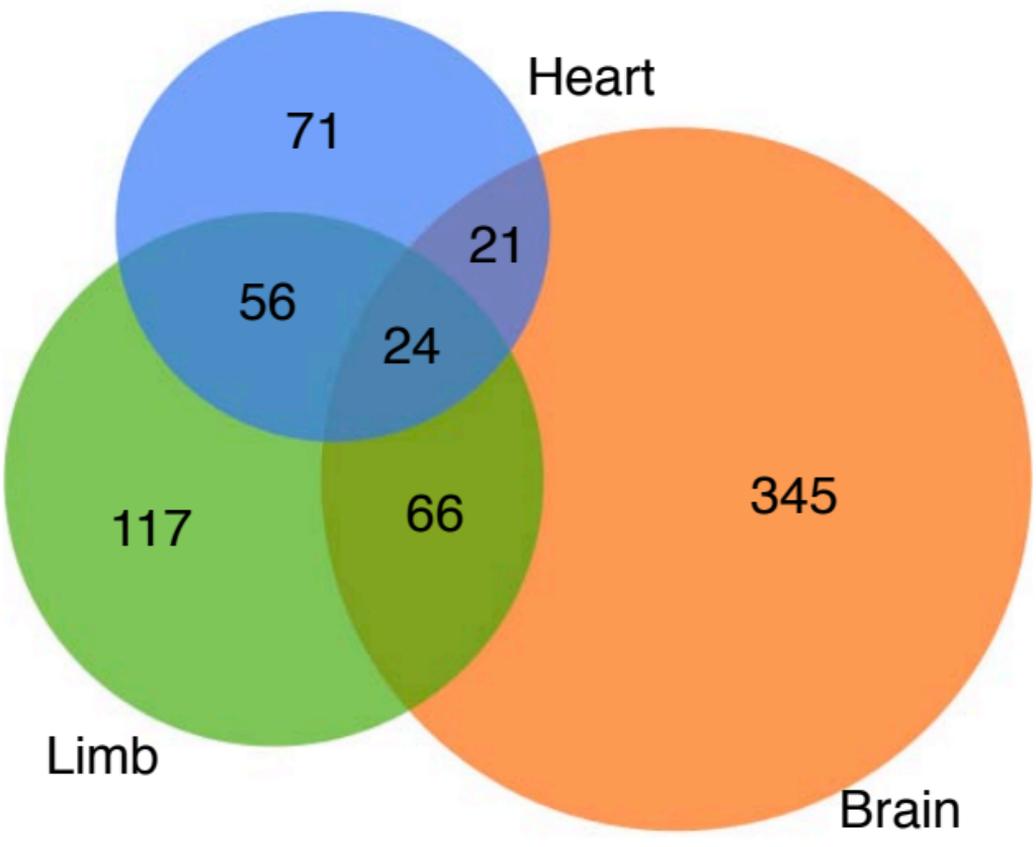

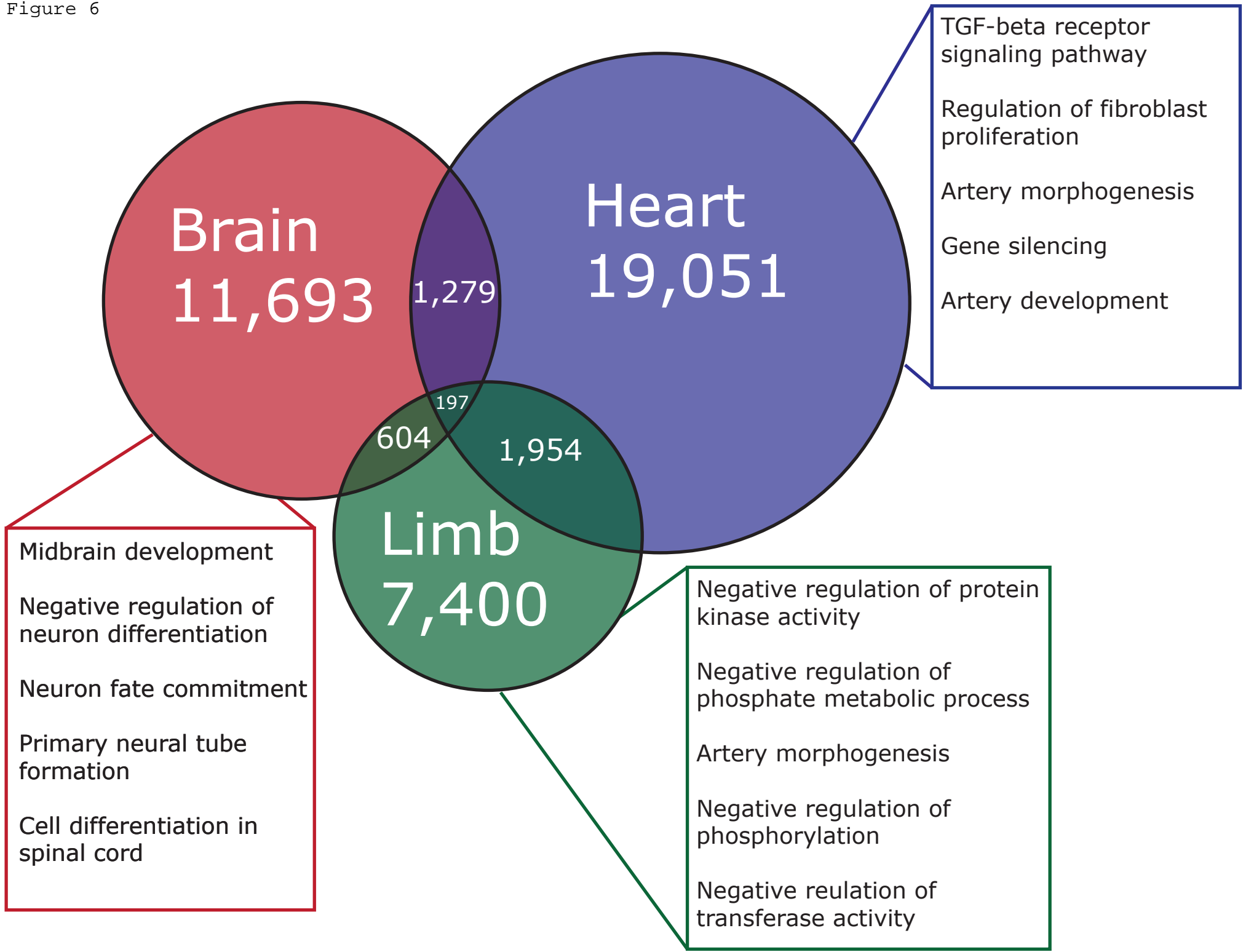

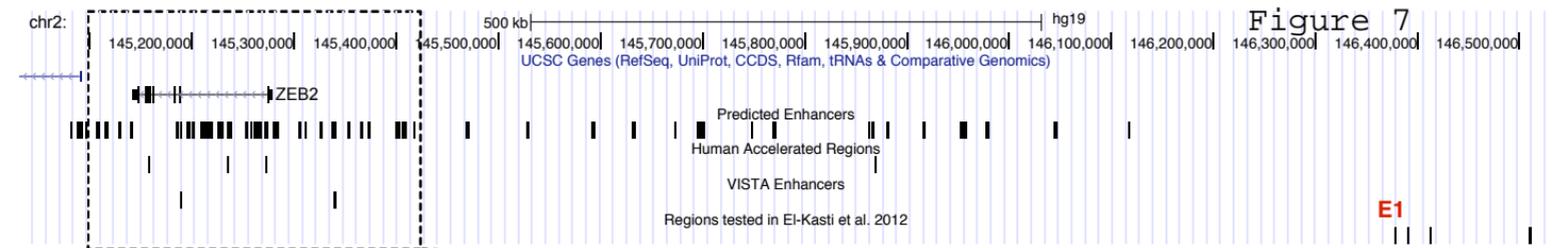

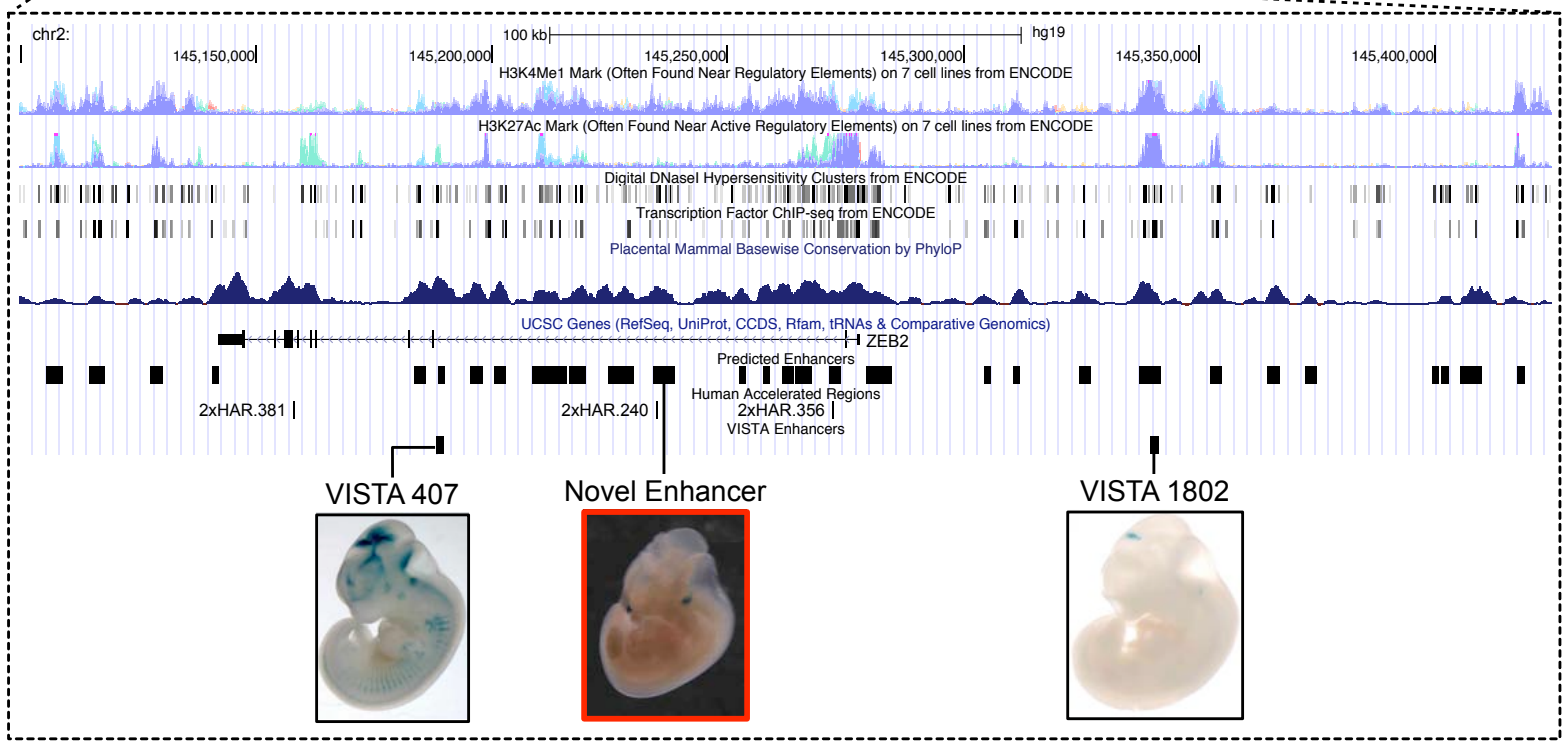

VISTA 407

Novel Enhancer

VISTA 1802